\documentclass[aps,prl,twocolumn,groupedaddress,showpacs]{revtex4}

\usepackage{graphicx}
\usepackage{latexsym}
\usepackage{verbatim}

\begin{document}

\title{Collective excitation of Bose-Einstein condensates in the transition region between 3D and 1D}

\author{M. Kottke$^1$}
\author{T. Schulte$^1$}
\email[]{schulte@iqo.uni-hannover.de}
\author{L. Cacciapuoti$^2$}
\author{D. Hellweg$^1$}
\author{S. Drenkelforth$^1$}
\author{W. Ertmer$^1$}
\author{J. J. Arlt$^1$}

\affiliation{$^1$ Institut f\"ur Quantenoptik, Universit\"at
Hannover, Welfengarten 1, 30167 Hannover, Germany\\$^2$ ESA Research
and Scientific Support Department, ESTEC, Keplerlaan 1 - P.O. Box
299, 2200 AG Noordwijk ZH, The Netherlands}

\date{\today}

\begin{abstract}

We measure the frequency of the low $m$=0 quadrupolar excitation
mode of weakly interacting Bose-Einstein condensates in the
transition region from the 3D to the 1D mean-field regime. Various
effects shifting the frequency of the mode are discussed. In
particular we take the dynamic coupling of the condensate with the
thermal component at finite temperature into account using a
time-dependent Hartree-Fock-Bogoliubov treatment developed in
\cite{giorgini}. We show that the frequency rises in the
transition from 3D to 1D, in good agreement with the theoretical
prediction~\cite{Menotti}.

\end{abstract}


\maketitle

One-dimensional (1D) quantum degenerate Bose gases have recently
attracted considerable theoretical and experimental
interest~\cite{QGLD}. On the one hand this interest is due to their
remarkable physical properties which are absent in three-dimensional
systems. On the other hand a rapid advance in trapping techniques
for ultracold gases has put these systems within experimental reach.
In particular optical lattices~\cite{bectalltice}, optical dipole
traps~\cite{grimmtrap}, and atom chips~\cite{chipsprospect} have
recently been used to realize such low-dimensional systems. For
further experiments under these conditions a good understanding of
the transition between the 3D and the 1D regime is therefore of
crucial importance. In this paper this transition region is
characterized experimentally by monitoring the oscillation of a
quantum degenerate Bose gas.

From a fundamental point of view, one of the most striking
features of 1D quantum degenerate Bose gases is the predominant
role played by quantum fluctuations. For spatially homogeneous 1D
systems, fluctuations of the phase rule out the existence of any
off-diagonal long-range order (ODLRO)~\cite{Kane} even at
temperature $T=0$ \cite{Schwartz}. The finite size of trapped 1D
gases however allows for a rich variety of possible scenarios,
including true phase-coherent Bose-Einstein condensates (BECs) as
well as phase-fluctuating BECs, the so-called
quasi-condensates~\cite{Petrov}.

The behavior of these 1D Bose gases is governed by the ratio of
interaction and kinetic energy
\begin{equation}
\gamma= m\,g_{1D}/\hbar^{2}\,n_{1D} \label{gamma}
\end{equation}
where $n_{1D}$ is the 1D atomic density, $m$ the atomic mass and
$g_{1D}$ the 1D coupling constant. Since this ratio scales as $1/n$
these gases counter-intuitively become more non-ideal when the
density is decreased. The fascinating features of these systems have
led to continued theoretical interest over the past decades. For
homogeneous 1D Bose gases with short-range interactions, the ground
state~\cite{Lieb1}, excitation spectrum~\cite{Lieb2} and
thermodynamic properties~\cite{Yang} of the system can be determined
with a Bethe Ansatz for arbitrary values of $ \gamma $. In the case
of trapped systems the equation of state can be found by combining
this approach with the local density approximation~\cite{Dunjko}.
For high densities the system is in the weakly interacting regime,
where it can be well described in the frame of mean-field theories
and ODLRO is present. On the contrary, for low densities the system
enters the strongly interacting or strongly correlated regime, where
a description by mean-field theories fails and ODLRO is strongly
reduced. In the limit $\gamma \rightarrow \infty $ the system is
equivalent to a 1D gas of impenetrable bosons, the so-called
Tonks-Girardeau (TG) gas~\cite{Girardeau}, where the bosonic
particles effectively acquire fermionic properties.

To experimentally prepare a 1D gas in a harmonic trap with
cylindrical symmetry one has to fulfill the condition
\begin{equation}
  \mu\,,k_{B}\,T\ll \hbar \,\omega_{\perp},
\label{1d}
\end{equation}
where $\omega_{\perp}$ denotes the radial trap frequency and $\mu$
is the chemical potential of the ensemble. If this condition is
fulfilled, neither the thermal nor the interaction energy is
sufficient to affect the radial shape of the ground-state wave
function. The radial degree of freedom is then frozen, since the
atoms are confined to the ground state of the radial trapping
potential.

The first experimental realization of a weakly interacting 1D BEC
was achieved in a magnetic trap with very high aspect ratio and
detected by a change in the ballistic
expansion~\cite{Goerlitz,Schreck}. The tight radial confinement
required to fulfill (\ref{1d}) can also be provided by magnetic
microtraps~\cite{1Dchip} or optical wave guides~\cite{waveguide}.
Such waveguides have allowed for the first realization of stable
matter-wave bright solitons~\cite{Solitons}. Two-dimensional optical
lattices~\cite{2dlattice} offer the possibility to realize arrays of
1D gases which are coupled via tunneling between the lattice sites.
In such a lattice gas a theoretically predicted reduction of the
three-body recombination rate within the strongly correlated
regime~\cite{Gangardt} has been experimentally
observed~\cite{Tolra}. Recently, seminal experiments have reached
the TG regime in 2D optical lattices~\cite{Paredes,Weiss}.

In this paper, we experimentally study collective excitations of
weakly interacting BECs in the transition region between 3D and
1D, i.e. for values of the chemical potential
$\mu\sim\hbar\,\omega_{\perp}$. The experimental investigation of
collective excitations allows for quantitative tests of the
underlying theoretical description. We focus on the observation of
the low $m=0$ quadrupolar mode. Collective modes were investigated
experimentally and theoretically in 3D
BECs~\cite{QPO3D1,QPO3D2,QPO3D5,QPO3D6,QPO3D7,Stringarimode,Edwards}
and in arrays of 1D gases confined in a 2D optical
lattice~\cite{Pedri,Moritz}. Our experimental investigation in the
transition region reveals clear signatures of one-dimensional
behavior and is complementary to experiments on the
phase-coherence properties~\cite{Dettmer} of the system.

The dynamics of the low $m$=0 mode consists of an out-of-phase
oscillation of the axial and radial diameters of the cloud. For
Bose gases confined in cylindrically symmetric traps the
hydrodynamic equations of superfluids~\cite{Pines} allow for the
analytic determination of the mode frequency at $T=0$: For the
case of a very elongated but 3D weakly interacting BEC the
frequency is $\omega=\sqrt{5/2} \,\omega_{z}$~\cite{stringari2},
where $\omega_{z}$ is the axial trap frequency. In the transition
from the 3D mean-field regime to the 1D mean-field regime, the
radial dynamics of the condensate freezes out, and accordingly the
radial component of the low $m$=0 mode vanishes. In this process,
the mode frequency rises to
$\omega=\sqrt{3}\,\omega_{z}$~\cite{Tin-Lun,stringari2}. For the
TG gas the frequency increases to $\omega= 2\,
\omega_{z}$~\cite{TGfreq}, identical to a non-interacting Fermi
gas. The frequency in the intermediate regimes has been
numerically determined by using a combination of hydrodynamic
equations with a sum-rule approach~\cite{Menotti} and later by
exclusively relying on hydrodynamic equations~\cite{Fuchs}. Note
that the analytic solutions given above depend on the axial trap
frequency only, whereas the mode frequency in the intermediate
regimes depends on the radial confinement as well~\cite{Menotti}.

\begin{figure}[h]
\includegraphics*[width=7cm]{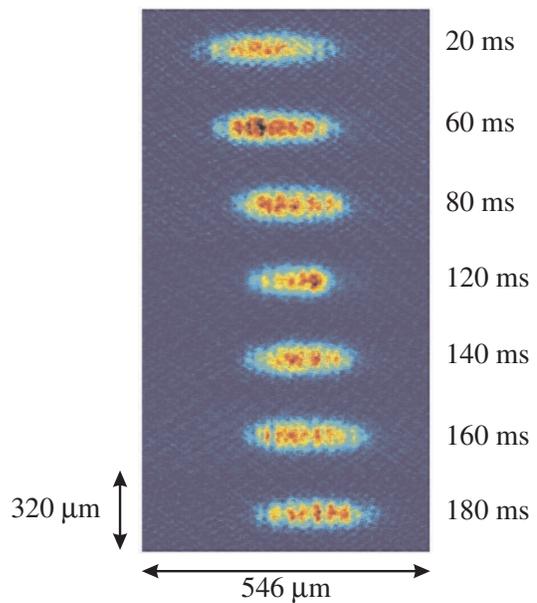}
\caption{Absorption images after $30 \,$ms time of flight for
various hold times $\tau$ in the magnetic trap (indicated on the
right). The oscillation of the quadrupolar mode is clearly visible.
The axial displacement of the cloud is due to the simultaneous
excitation of the dipole mode. Axial density modulations observable
in the absorption images are related to the presence of phase
fluctuations in the condensate~\cite{Dettmer}. We have vertically
compressed the images for graphical representation. \label{Figure1}}
\end{figure}

Our experiments were performed with $^{87}$Rb Bose-Einstein
condensates in the $|F\!=\!1, m_{F}\!=\!-1\rangle $ hyperfine
ground state confined in a strongly elongated magnetic trap. The
axial trap frequency was $3.40 \,$Hz and the radial trap frequency
was varied between $265 \,$Hz and $385 \,$Hz, resulting in aspect
ratios as low as $\lambda = \omega_z/\omega_{\perp} \approx
1/113$. Further details of our experimental apparatus can be found
elsewhere~\cite{Hellweg}. In a trap with fixed geometry, the 1D
criterion (\ref{1d}) reduces to a condition on the number of atoms
in the sample. Neglecting the contribution of thermal atoms to the
chemical potential, the gas is 1D if the number of condensed atoms
$N_{0}$ is smaller than~\cite{Goerlitz}
\begin{equation}
N_{0}^{1\mathrm{D}}=\sqrt{\frac{32}{225}}\,
\frac{a_{z}^2}{a\,a_{\perp}}, \label{N1d}
\end{equation}
where $a_{\perp}$ and $a_{z}$ denote the harmonic oscillator
length in the radial and axial directions and $a$ is the s-wave
scattering length. For our trap parameters
$N_{0}^{1\mathrm{D}}\approx 4000$. To approach this regime we
reduce the number of atoms by adjusting the end of the forced
evaporation ramp. We are restricted by our detection system to a
minimum number of $N \approx 2000$ atoms. For this minimum number
of atoms $\mu/\hbar \omega_{\perp} \approx 0.75$, thus we obtain
ensembles on the border of one-dimensionality.

\begin{figure}
\includegraphics*[width=8.6cm]{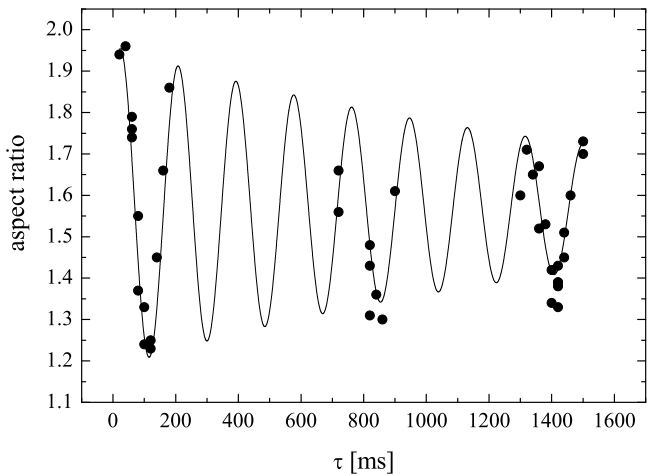}
\caption{Aspect ratio of the clouds after time of flight as a
function of the hold time in the magnetic trap. The fit to the
data gives a mode frequency of $\omega=5.41\pm0.02 \,$Hz. The data
points originate from clouds with an average dimensionality
parameter $P=13.3$ and reduced temperature $T/T_{c}=0.34$.
\label{Figure2}}
\end{figure}

We performed the following experimental procedure: Laser-cooled
atoms were loaded into a moderately elongated magnetic trap with
$\lambda \approx 1/25$. Radio-frequency evaporative cooling was
performed to obtain temperatures just above the transition
temperature. Then the trap was axially decompressed to reach the
desired aspect ratio, and the final evaporation ramp to obtain BEC
was performed in the strongly elongated trap. The low $m$=0 mode
was excited by modulating the current of the magnetic trap for
five periods at a frequency close to the expected value of the
mode frequency $\omega$. The condensate was allowed to oscillate
in the magnetic trap for a hold time $\tau$. Subsequently, the
trapping potential was switched off and the atomic density
distribution was detected by resonant absorption imaging after $30
\,$ms time of flight. A bimodal fit to the density distribution
was used to extract the aspect ratio, the total and condensed
particle numbers and the temperature of the ensemble. By varying
the hold time $\tau$ in the magnetic trap the oscillation was
stroboscopically monitored as shown in Fig.~\ref{Figure1}. Since
the mode frequency depends on the dimensionality of the system,
the data was sorted according to the parameter $P=N_{0}\lambda
a/a_{\perp}$, which appropriately describes the degree of
dimensionality for our purposes~\cite{Menotti}.
Figure~\ref{Figure2} shows the aspect ratio of clouds within an
interval of the parameter $P$ versus the hold time $\tau$. To
extract the frequency we fit this data with a damped sinusoidal
function.

The frequency rise which accompanies the transition from the 3D to
the 1D mean-field regime constitutes a shift of only $10 \%$. Due
to the small size of this effect other influences on the
oscillation frequency have to be considered carefully. In
principle, frequency shifts can be caused by effects beyond the
mean-field approximation, by large oscillation amplitudes and by
the finite temperature of the system.

The frequency shift due to corrections to the mean-field
approximation can be analyzed in the frame of the hydrodynamic
theory of superfluids. Using the first quantum correction to the
Bogoliubov equation of state, this shift of the lowest quadrupolar
mode at $T=0$ is~\cite{BeyondMF}
\begin{equation}
\frac{\delta \omega}{\omega}=\frac{63
\sqrt{\pi}}{128}\sqrt{a^{3}n(0)}f(\lambda), \label{depletion}
\end{equation}
where $f(\lambda)$ is a function that depends on the geometry of
the trap (see Eqn. (17) in~\cite{BeyondMF}). Since the gas
parameter $a^{3}n(0)$ is very small for our experimental
conditions, this frequency shift is below the $0.1 \% $ level and
thus negligibly small.

The hydrodynamic equations are also well suited to investigate the
effect of large oscillation amplitudes at $T=0$. Large amplitudes
cause nonlinear coupling between the normal modes of the system and
shifts of the oscillation frequency. The frequency shift for the low
$m=0$ quadrupolar mode is~\cite{LargeAm}
\begin{equation}
\frac{\delta \omega(A_z)}{\omega} =\frac{ \delta(\lambda)
A_z^2}{16},
\end{equation}
where $A_{z}$ is the relative oscillation amplitude of the
condensate length in the trap and $\delta(\lambda)$ is a factor
depending on the trap geometry (see Eqn. (22) in~\cite{LargeAm}).
The amplitude $A_{z}$ can be extracted from the oscillation
amplitude in the time of flight images $A_z^{\mathrm{TOF}}$ by
using scaling theory~\cite{scaling}. For our parameters we obtain
$A_z \approx A_z^{\mathrm{TOF}} / \sqrt{2}$ and the oscillation
amplitudes in the trap are $A_z \leq 20\,$\% for our measurements.
Hence the corresponding frequency shift is limited to $\delta
\omega (A_z)/\omega \leq 0.5\,$\% and is thus clearly smaller than
the expected shift due to dimensional effects.

Let us now turn to the most important correction to the mode
frequency, which is caused by the finite temperature of the
system. Early experiments on collective excitations in 3D systems
investigated effects due to the finite temperature of the sample,
including frequency shifts with respect to the zero temperature
predictions~\cite{finiteTJila,finiteTMIT}. On the theory side
various approaches have been proposed to include finite
temperature into the theory of weakly interacting trapped gases.
To describe the observed features, such as the damping of
excitations or the dynamic coupling between the condensed and
thermal components, time-dependent mean-field schemes are
appropriate. In this paper, we rely on the linearized
time-dependent Hartree-Fock-Bogoliubov approach proposed
in~\cite{giorgini}. Within this approach the coupled equations for
the dynamics of the condensate and the thermal component are not
solved self-consistently, but perturbatively up to second order in
the coupling constant $g$. Thus the non-physical gap in the
self-consistent static quasi-particle excitation spectrum is
replaced by a well-behaved gapless Bogoliubov type of spectrum.
Moreover at $T=0$ this approach recovers the first order quantum
correction (\ref{depletion}), and therefore includes the frequency
shift due to corrections to the mean-field approximation.

\begin{figure}[h]
\includegraphics*[width=8.6cm]{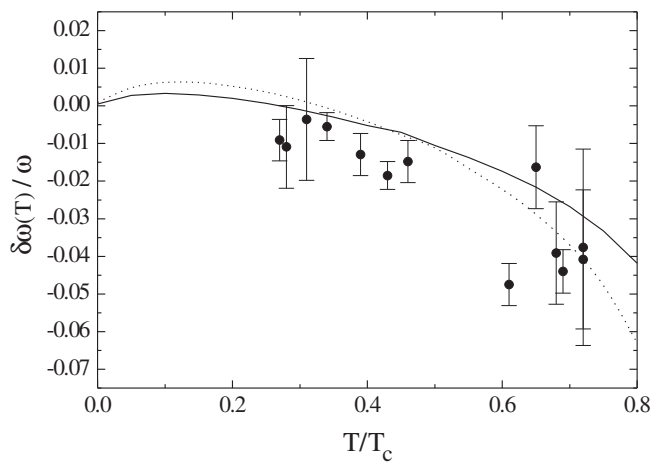}
\caption{Normalized frequency difference between our measurements
and the $T$=0 prediction in \cite{Menotti} as a function of the
reduced temperature. The lines indicate the theoretical frequency
corrections for an interaction parameter of $\eta=0.33$ (solid)
and $\eta=0.41$ (dotted) according to \cite{giorgini}.
\label{Figure3}}
\end{figure}

Figure~\ref{Figure3} shows a comparison between the measured
frequency shifts with respect to the $T=0$ calculation
in~\cite{Menotti} and the theoretical prediction for the shifts
due to finite temperature according to~\cite{giorgini}. The
calculated frequency shift depends on temperature and on the
parameter $\eta =\mu/\left(k_{B}\,T_{c}\right)$ which describes
the interactions in a 3D Bose gas~\cite{giorgini}. Here $T_{c}$
denotes the critical temperature for a non-interacting gas. Due to
the variation of the atom number in the experiments, the
corresponding values for $\eta$ range from 0.33 to 0.41. The
theoretical prediction for these cases is shown in
Fig.~\ref{Figure3}. Despite the low atom numbers used in some of
the measurements this theory, valid in the thermodynamic limit,
shows good agreement with the measured frequency shifts. Since the
finite-temperature shifts are of the order of a few percent, they
constitute an important contribution to the mode frequency. For a
subsequent correction of the mode frequency, the parameter $\eta$
and the resulting frequency shift was calculated for each data
point.

\begin{figure}
\includegraphics*[width=8.6cm]{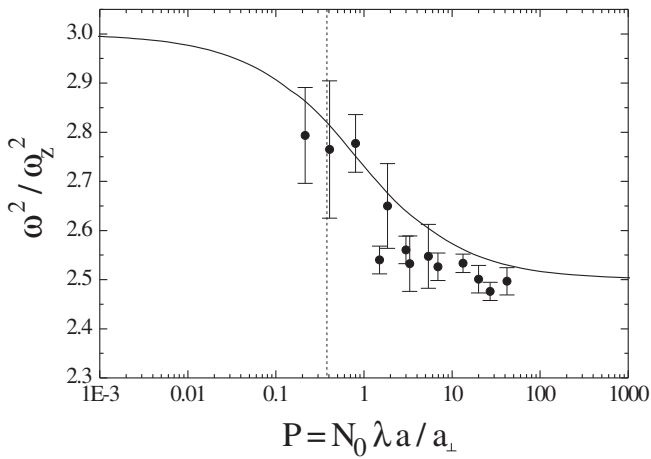}
\caption{Measurement of the mode frequency $\omega$ for the
transition between the 3D and the 1D mean-field regimes. The
frequency shifts due to finite temperature and large oscillation
amplitudes are included. The solid line shows the theoretical
prediction \cite{Menotti}. The dashed vertical line indicates the
transition point between the two regimes according to (\ref{N1d}).
\label{Figure4}}
\end{figure}

To compare our finite-temperature measurements with the
zero-temperature prediction in~\cite{Menotti}, we correct our data
for the finite-temperature shift and the smaller shift due to the
oscillation amplitude. Figure~\ref{Figure4} shows the measured
mode frequencies, including these corrections. The variation in
the number of particles allowed for measurements covering almost
the entire transition region between the 3D and 1D mean-field
regimes. Within the experimental error, our data shows good
agreement with the predicted frequency dependence at zero
temperature. The error increases towards small values of the
dimensionality parameter $P$ due to the small particle numbers
involved in these measurements. A small discrepancy between the
theoretical prediction and the measurement might be due to the
loss of atoms during the oscillation or systematic uncertainties
in the determination of the number of atoms.

In conclusion, we have measured the increase of the oscillation
frequency of the low $m$=0 quadrupolar mode in the crossover region
between the 3D and 1D regimes. To compare our measurements with the
zero-temperature prediction, various effects that can lead to shifts
of this frequency were carefully evaluated. We have identified
finite-temperature effects as an important contribution to the
frequency shift. Including frequency corrections due to the finite
temperature and the oscillation amplitude, our data shows good
agreement with the theoretical prediction. The observed frequency
increase constitutes a clear signature of the onset of
one-dimensionality. These results confirm that mode frequency
measurements provide a sensitive probe of the dimensionality of
quantum degenerate Bose gases.

We thank C. Menotti and S. Stringari for providing the numerical
data of Fig.~\ref{Figure4} and acknowledge fruitful discussion with
S. Giorgini and M. Lewenstein. This work was supported by the {\it
Deutsche Forschungsgemeinschaft} within the SFB\,407.

\end{document}